\documentclass[%
 reprint,
superscriptaddress,
nofootinbib,
 amsmath,amssymb,
 aps
]{revtex4-1}

\usepackage{color,graphicx,epsfig}
\usepackage{ifpdf}
\usepackage{amsmath}
\usepackage{bm}
\usepackage{color}
\usepackage[english]{babel}
\usepackage{graphicx}%
\usepackage{amsfonts}%
\usepackage{amssymb}
\usepackage{braket}
\usepackage{hyperref}
\usepackage{enumerate}

\definecolor{nicered}{rgb}{0.7,0.1,0.1}
\definecolor{nicegreen}{rgb}{0.1,0.5,0.1}
\hypersetup{colorlinks,citecolor= nicegreen,linkcolor= nicered}

\newcommand{\be}{\begin{equation}}
\newcommand{\ee}{\end{equation}}
\newcommand{\bea}{\begin{eqnarray}}
\newcommand{\eea}{\end{eqnarray}}

\definecolor{Red}{rgb}{1.,0.,0.}

\newcommand{\sign}{\text{sgn}}

\def\OMIT#1{}

\begin{document}

\def\Argonne{High Energy Physics Division, Argonne National Laboratory, Argonne, IL 60439, USA}
\def\Northwestern{Department of Physics \& Astronomy, Northwestern University, Evanston, IL 60208, USA}


\author{Radja Boughezal}
\email[Electronic address: ]{rboughezal@anl.gov}
\affiliation{\Argonne}

\author{Hai Tao Li}
\email[Electronic address: ]{haitao.li@northwestern.edu}
\affiliation{\Argonne}
\affiliation{\Northwestern}

\author{Frank Petriello}
\email[Electronic address: ]{f-petriello@northwestern.edu}
\affiliation{\Argonne}
\affiliation{\Northwestern}

\title{$W$-boson production in polarized proton-proton collisions at RHIC through next-to-next-to-leading order in perturbative QCD}

\date{\today}
\begin{abstract}
We perform a study of $W$-boson production in polarized proton-proton collisions through next-to-next-to-leading order (NNLO) in perturbative QCD. This calculation is required to extend the extraction of polarized parton distribution functions to NNLO accuracy. We present differential distributions at $\sqrt{s}=510$ GeV, relevant for comparison to measurements from the Relativistic Heavy Ion Collider (RHIC). The NNLO QCD corrections significantly reduce the scale dependence of the cross section. We compare the longitudinal single-spin asymmetries as a function of lepton pseudorapidity to RHIC data. The asymmetries exhibit excellent stability under perturbative QCD corrections.
\end{abstract}

\maketitle

\section{Introduction} \label{sec:intro}

An understanding of how the spin of the proton arises from its partonic constituents is an outstanding 
problem in fundamental physics.  It is one of the main science drivers of the future Electron-Ion Collider~\cite{Accardi:2012qut,Aschenauer:2014cki}.  Over the past several 
years new insight into aspects of polarized proton structure has been provided by the Relativistic Heavy Ion Collider (RHIC) at Brookhaven 
National Laboratory.  Jet and pion production in polarized proton-proton collisions at RHIC indicate that gluons contribute a non-negligible 
amount to the proton spin~\cite{deFlorian:2014yva}.  Another recent result which motivates this paper is the first evidence of flavor asymmetry in the polarized sea. The relevant observables that lead to this result are the single-spin asymmetries in polarized $pp \to W^{\pm}$, where the $W$-bosons 
decay via $W^+ \to e^+ \bar{\nu}$ and $W^- \to e^- \nu$: $A_L = (\sigma^+ - \sigma^-)/(\sigma^+ + \sigma^-)$, where $\sigma^{\pm}$ denotes 
the cross section with a single proton beam with positive or negative helicity. 
 The asymmetries were proposed a long time ago~\cite{Bourrely:1993dd,Bourrely:1994sc} to probe the flavor structure of polarized parton distributions. 
At leading-order in QCD perturbation theory, the asymmetries 
are proportional to the following combinations of parton distribution functions:
\bea
A_L^{W^+} \propto \frac{\Delta \bar{d}(x_1)u(x_2) - \Delta u(x_1) \bar{d}(x_2)}{\bar{d}(x_1)u(x_2) + u(x_1) \bar{d}(x_2)}, \nonumber \\
A_L^{W^-} \propto \frac{\Delta \bar{u}(x_1)d(x_2) - \Delta d(x_1) \bar{u}(x_2)}{\bar{u}(x_1)d(x_2) + d(x_1) \bar{u}(x_2)},
\eea
where $x_{1,2}$ are the Bjorken-$x$ momentum fractions.  They are related to the $W$-boson mass $M_W$ and rapidity $Y_W$ according to $x_{1,2} = (M_W /\sqrt{s}) e^{\pm Y_W}$, where $\sqrt{s}$ is the collider center-of-mass energy. The functions $u(x),d(x)$ and $\bar{u}(x),\bar{d}(x)$ are respectively the valence and sea quark parton distribution functions (PDFs) of the proton.  
The functions $\Delta u(x),\Delta d(x)$ and $\Delta\bar{u}(x),\Delta \bar{d}(x)$ are the corresponding polarized PDFs. The polarized and unpolarized PDFs can be written as 
\bea
    f_{a/N} =&\frac{1}{2}\sum_{s_a,s_N} f_{a/N}^{s_a/s_N}\;, 
    \nonumber \\
    \Delta f_{a/N} =&\frac{1}{2} \sum_{s_a,s_N} \sign_{s_a\times s_N} f_{a/N}^{s_a/s_N}\;,
\eea
where $s_a$ and $s_N$ are the helicities of parton $a$ and nucleon $N$. For a recent review of unpolarized and polarized PDFs we refer the reader to Ref.~\cite{Ethier:2020way}.
Assuming the dominance of valence distributions as the Bjorken-$x$ values approach one, we find $A_L^{W^+} \to - \Delta u(x_1) /u(x_1)$ and $A_L^{W^-} \to - \Delta d(x_1) /d(x_1)$ for large positive rapidity where $x_1 \gg x_2$, while for large negative rapidity where $x_2 \gg x_1$ we find $A_L^{W^+} \to  \Delta \bar{d}(x_1) /d(x_1)$ and 
$A_L^{W^+} \to  \Delta \bar{u}(x_1) /u(x_1)$. For negative $Y_W$ these asymmetries therefore give access to the polarized quark sea.  
Results from the PHENIX and STAR experiments at RHIC indicate a significant non-zero $\Delta \bar{u}(x)-\Delta \bar{d}(x)$ in the region $0.05 < x <0.25$~\cite{Aggarwal:2010vc,Adamczyk:2014xyw,Adare:2018csm,Adam:2018bam}.

Measurements such as the one above are typically incorporated into global analyses of proton structure such that information from all 
relevant experiments can inform the obtained PDFs.  These global analyses are performed at a given order in perturbative QCD, which indicates to what order the evolution kernels and hard-scattering cross sections are calculated.  The PDFs of an unpolarized proton are determined to next-to-next-to-leading order (NNLO) in several global analyses~\cite{Harland-Lang:2017ytb,Ball:2017nwa,Alekhin:2017kpj,Hou:2019efy,Butterworth:2015oua}.  The polarized PDFs are currently only known to the NLO~\cite{Nocera:2014gqa,deFlorian:2014yva,Ethier:2017zbq}. It is expected that a much sharper picture of polarized proton structure will be obtained at an EIC, and that the precision of the data will warrant a NNLO extraction of polarized PDFs like in the unpolarized case. One necessary component of this extension of the polarized PDF extraction to NNLO is the calculation of the requisite hard scattering cross sections to NNLO in perturbative QCD.

In this paper we take a step toward this goal by calculating $W$-boson production in polarized proton-proton collisions to NNLO in perturbative QCD.  Many efforts have previously been devoted to studying the NLO QCD corrections~\cite{Nadolsky:2003fz,Nadolsky:2003ga,deFlorian:2010aa,Ringer:2015oaa}. We utilize the $N$-jettiness subtraction approach first developed for unpolarized proton-proton collisions at the Large Hadron Collider~\cite{Boughezal:2015dva,Gaunt:2015pea}. Our result represents its first application to polarized collisions, and the first prediction for a cross-section at NNLO in polarized proton-proton collisions.  We discuss aspects of the phenomenology of polarized $W$-boson production at RHIC energies, and perform a comparison to the existing STAR data.  The NNLO perturbative corrections exhibit excellent convergence, and their inclusion 
improves the residual scale uncertainty often used to estimate the theoretical errors.  

This paper is organized as follows. In Section~\ref{sec:setup} we introduce the formalism for our calculation and review the $N$-jettiness subtraction method. In Section~\ref{sec:numerics} we present our results for  several differential distributions and the single-spin asymmetry. We conclude in Section~\ref{sec:conc}. 

\section{Calculational framework} \label{sec:setup}

We review here the theoretical formalism used in our calculation.  The collinear factorization theorem of perturbative QCD allows us to 
write the cross section for polarized proton-proton collisions as
\be
\sigma^{s_1 s_2} = \sum_{a,b} \sum_{s_a,s_b} f^{s_a/s_1}_{a/N_1} \otimes f^{s_b/s_2}_{b/N_2} \otimes \hat{\sigma}^{s_a,s_b}_{a,b}.
\ee 
Here $s_{1,2}$ refer to the helicities of nucleons $N_{1,2}$, $s_{a,b}$ are the helicities of the partons that enter the hard scattering process, 
$f^{s_a/s_1}_{a/N_1}$ denotes the PDF for taking parton $a$ with helicity $s_a$ out of a nucleon with helicity $s_1$, and $\hat{\sigma}^{s_a,s_b}_{a,b}$ is the partonic hard-scattering process.  We have schematically denoted the integration over Bjorken-$x$ values with the symbol $\otimes$. Two combinations of helicities are needed to describe polarized collisions at RHIC, namely the unpolarized cross section and the polarized results arising from a single beam polarized:
\bea
\sigma &=& \frac{1}{4} \left( \sigma^{++}+\sigma^{+-}+\sigma^{-+}+\sigma^{--}\right), \nonumber \\
\Delta\sigma &=& \frac{1}{4} \left( \sigma^{++}+\sigma^{+-}-\sigma^{-+}-\sigma^{--}\right).
\eea
Using the parity conservation of QCD we can write the unpolarized and polarized cross sections as
\bea
\sigma &=& \sum_{a,b} f_{a/N_1} \otimes f_{b/N_2} \otimes\hat{\sigma}_{ab}, \nonumber \\
\Delta\sigma &=& \sum_{a,b} \Delta f_{a/N_1} \otimes f_{b/N_2} \otimes\Delta \hat{\sigma}_{ab},
\eea
where $f, \Delta f$ respectively denote the standard unpolarized and polarized PDFs.  The polarized partonic cross section is
\be
\Delta \hat{\sigma} = \frac{1}{4}  \left( \hat{\sigma}^{++}+\hat{\sigma}^{+-}-\hat{\sigma}^{-+}-\hat{\sigma}^{--}\right).
\ee
The single-spin asymmetry defined in Section~\ref{sec:intro} can be written as $A_L = \Delta \sigma / \sigma$.

To facilitate the calculation of the NNLO corrections to the partonic hard-scattering cross section it is convenient to study the cross section in the limit of small zero-jettiness ($\tau_0$), a hadronic event-shape variable that has been extensively discussed in the literature~\cite{Stewart:2010tn}. 
The definition of $\tau_0$ reads
\begin{align}
    \tau_0 = \sum_{k} \min\left\{{n\cdot p_k, \bar{n} \cdot p_k } \right\}\;,
\end{align}
where $n=(1,\hat{z})$ and $\bar{n} = (1,-\hat{z})$ with beam direction $\hat{z}$. We will use $\tau_0$ defined in the $W$-boson rest frame, leading to the so-called leptonic definition of the N-jettiness variables.

The $\tau_0$ variable encapsulates all of the singular behavior associated with additional radiation in color-singlet $W$-boson production. 
In the small-$\tau_0$ limit the unpolarized and polarized cross sections can be factorized as~\cite{Stewart:2009yx,Stewart:2010pd}
\bea \label{eq:below}
\frac{d\sigma}{d\tau_0} = \sum_{a,b} H_{ab} \times B_{a/N_1} \otimes B_{b/N_2} \otimes S_{ab} [1+\mathcal{O}(\tau_0)], \nonumber \\
\frac{d\Delta\sigma} {d\tau_0}= \sum_{a,b} \Delta H_{ab}\times \Delta B_{a/N_1} \otimes B_{b/N_2} \otimes S_{ab}  [1+\mathcal{O}(\tau_0)]. \nonumber \\
\label{eq:facthm}
\eea
Here, the $B$ and $\Delta B$ denote the beam functions that describe collinear radiation from unpolarized and polarized initial-state nucleons respectively, the soft function $S$ describes soft radiation, and $H$ and $\Delta H$ are the hard functions which encode the virtual corrections.  The symbol $\otimes$ denotes here the convolution over $\tau_0$ arising from each type of radiation. There are power corrections to these formulae that vanish in the $\tau_0 \to 0$ limit and that are explicitly denoted above. It is known from the existing literature~\cite{Stewart:2010tn,Stewart:2010pd} that the leptonic definition of $\tau_0$ reduces the impact of power corrections at the LHC. We also note that the power corrections that vanish in the $\tau_0 \to 0$ limit can be systematically calculated~\cite{Moult:2016fqy,Boughezal:2016zws,Moult:2017jsg,Boughezal:2018mvf, Ebert:2018lzn,Boughezal:2019ggi}.

 The hard and soft functions can be directly calculated in QCD perturbation theory, while the beam functions are non-perturbative objects.  Their infrared behavior is identical to that of the PDFs, and they can be matched onto these quantities:
\bea
B_{a/N_1} &=& \sum_b I_{ab} \otimes f_{b/N_1} , \nonumber \\
\Delta B_{a/N_1} &=& \sum_b \Delta I_{ab} \otimes \Delta f_{b/N_1},
\eea
where $I$ and $\Delta I$ are perturbative matching coefficients.  We can identify the partonic hard scattering cross sections in the small-$\tau_0$ limit in terms of the object that appear in the factorization theorem as
\bea
\hat{\sigma}_{ab} &=& \sum_{c,d} I_{ac} \otimes I_{bd} \otimes S_{ab} \times H_{ab}, \nonumber \\
\Delta \hat{\sigma}_{ab} &=& \sum_{c,d} \Delta I_{ac} \otimes I_{bd} \otimes S_{ab} \times \Delta H_{ab}.
\eea

The use of the zero-jettiness event-shape variable to facilitate NNLO calculations is through the use of $N$-jettiness subtraction~\cite{Boughezal:2015dva,Gaunt:2015pea}. The basic idea of $N$-jettiness subtraction is to separate the low-$\tau_0$ limit from finite values through the use of a cutoff parameter $\tau_{\rm cut}$.  In the low-$\tau_0$ region the cross section is calculated using the factorization theorem above in Eq.~(\ref{eq:below}), with the $I$, $S$, and $H$ functions calculated in fixed-order perturbation theory.  The beam function~\cite{Stewart:2010qs,Gaunt:2014xga} and soft function~\cite{Kelley:2011ng,Monni:2011gb,Hornig:2011iu} are known to the requisite NNLO. Since $\tau_0$ captures all singularities in color-singlet production, above $\tau_{\rm cut}$ the cross section becomes a NLO calculation for $W$+1-jet production, which can be calculated with standard techniques such as dipole subtraction~\cite{Catani:1996vz}. 
The polarized and unpolarized cross sections can be written as 
\bea
    \sigma = \int_0^{\tau_\text{cut}}d\tau_0 \;\frac{ d\sigma}{d\tau_0} + \int_{\tau_\text{cut}}^{\tau_\text{max}}d\tau_0\; \frac{ d\sigma}{d\tau_0}\;.
    \label{eq:abovebelow}
\eea

The application of $N$-jettiness subtraction to unpolarized collisions and its implementation in efficient numerical codes, in particular for unpolarized $W$-boson production, has been discussed in detail in the literature~\cite{Boughezal:2016wmq}.  The primary numerical issue is the cancellation of logarithmic dependence on $\tau_{\rm cut}$ between the above-cut and below-cut contributions in Eq.~(\ref{eq:abovebelow}). The possibility of computing the power corrections to the below-cut term in the small-$\tau_0$ limit allows $\tau_{\rm cut}$ to be increased and the degree of logarithmic cancellation to be reduced, and the performance of this method can therefore be systematically improved~\cite{Moult:2016fqy,Boughezal:2016zws,Moult:2017jsg,Boughezal:2018mvf,Boughezal:2019ggi}. We focus here on the aspects that differ for polarized collisions.  As can be seen from the result of Eq.~(\ref{eq:facthm}) only the hard function and beam-function matching coefficients change in the polarized case.  The leading-order hard functions for the unpolarized and polarized cross sections are respectively just the leading-order partonic hard-scattering cross sections for the $q\bar{q} \to W^{\pm} \to l^{\pm} \nu$ process.  Helicity conservation along massless quark lines implies that the virtual corrections to the polarized hard function are identical to the unpolarized case, and can be obtained from QCD corrections to the vector-current form factor which is known to three loops~\cite{Baikov:2009bg,Gehrmann:2010ue}. Only the polarized matching coefficients require a new calculation, which is available in the literature~\cite{Boughezal:2017tdd}.  We incorporate our result into the MCFM~8.0 code framework~\cite{Boughezal:2016wmq}, which already contains an implementation of $N$-jettiness subtraction for color-singlet production in unpolarized collisions.

For the cross section above $\tau_{\rm cut}$ we make use of the helicity-dependent matrix elements for $W$+1-jet in MCFM~\cite{Campbell:2010ff}.  In the case that the first proton beam is polarized, we must replace the dipole subtraction terms if the first parton is the emitter, while keeping the other configurations the same as the unpolarized case. The dipole subtraction terms for polarized matrix elements can be found in Ref.~\cite{Czakon:2009ss} for arbitrary helicity eigenstates. Order-$\epsilon$ terms are required for integrated dipole subtraction terms where the LO polarized splitting functions are used~\cite{Vogelsang:1996im}.

\begin{figure}
    \centering
    \includegraphics[width=0.5\textwidth]{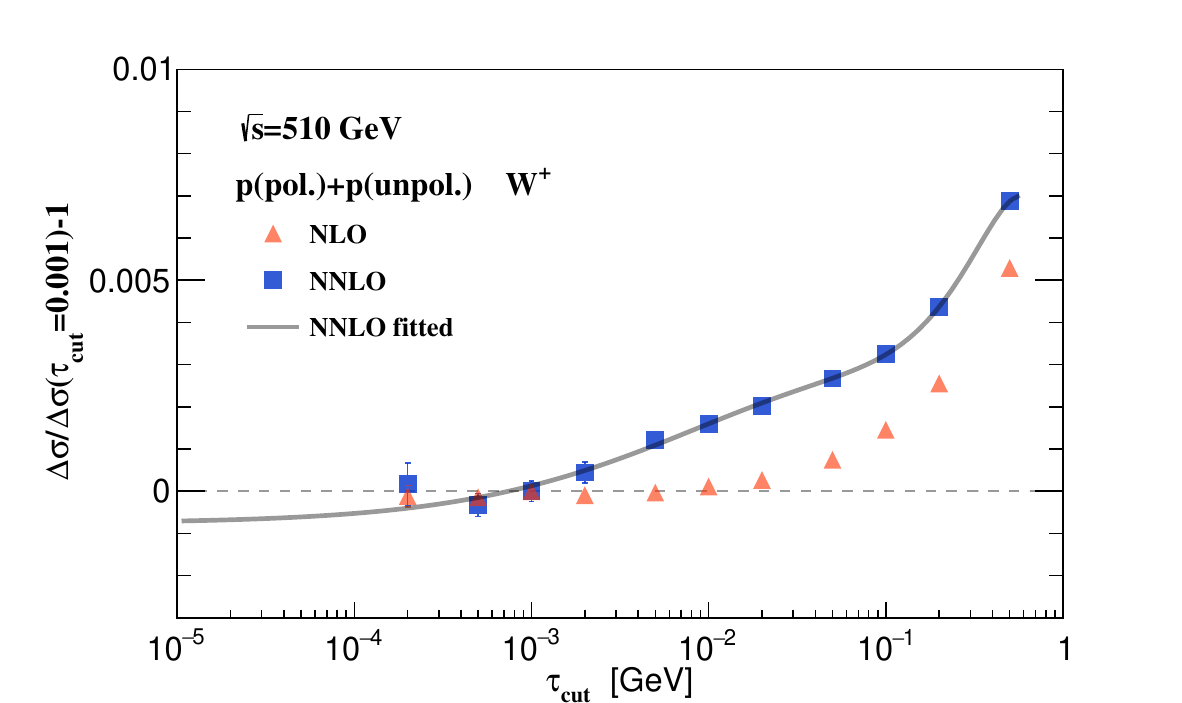}
    \vspace{-0.6cm}
    \caption{$\tau_{\rm cut}$ dependence of the NLO and NNLO polarized cross section for $W^+$ production. The y-axis denotes $\Delta \sigma/\Delta \sigma(\tau_{\rm cut}=0.001)-1$. The gray line represents the fitted $\tau_{\rm cut}$ dependence of the NNLO cross sections.  For most of the points the error bars are too small to be seen.  }
    \label{fig:taucut}
\end{figure}

To validate our $N$-jettiness subtraction approach for the polarized cross section, we first note that we have confirmed that the NLO cross section obtained using this technique matches exactly the result obtained using dipole subtraction in the small-$\tau_{\rm cut}$ limit.  In order to check our NNLO calculation we investigate the $\tau_{\rm cut}$ dependence of the inclusive cross section $\Delta\sigma$ at NLO and NNLO for $W^+$ production.  This is presented in Fig.~\ref{fig:taucut} with $\tau_{\rm cut}\in [0.0002,0.5]$. The $y$-axis is the relative difference to $\Delta\sigma$ with $\tau_{\rm cut}=0.001$. Clearly the logarithmic dependence on $\tau_{\rm cut}$ is cancelled between the below-cut and above-cut contributions, as is required when validating $N$-jettiness subtraction cross sections~\cite{Boughezal:2015dva,Gaunt:2015pea,Boughezal:2016wmq}.  The dependence of the total cross section on $\tau_{\rm cut}$ becomes negligibly small when the cutoff parameter is chosen sufficiently small. To check our results we perform a fit of the NNLO corrections to the functional form $ a_0 + x(a_3 \ln^3 x + a_2\ln^2 x + a_1 \ln x + b)$, following the approach in Ref.~\cite{Moult:2016fqy}. The result of the fit is consistent with this functional form, suggesting that all leading-logarithmic powers of $\tau_{\rm cut}$ are canceled between the above-cut and below-cut contributions to the cross section. The fitted NNLO correction to the cross section at $\tau_{\rm cut}=0$ is $\delta\Delta\sigma_{\rm NNLO}=-2694\pm 8$ fb. In the remainder of this paper we  choose  $\tau_{\rm cut}=0.001$ as default, where the cross section is $-2730\pm 12$ fb, consistent with the fitted corrections within 2.5$\sigma$. The difference between the two results is around 1\% of the NNLO correction, which can be taken as a measure of the systematic error in our calculation. We note that the NNLO correction is itself only an approximately 5\% correction to the total cross section, indicating that we have few-per-mille control over our result. We note that the NNLO power corrections to polarized cross sections are not yet known, and that a detailed derivation of them is beyond the scope of this work. 

\section{Numerical results} \label{sec:numerics}

We present here numerical results for  $W^{\pm}$ production in polarized $pp$ collisions at NNLO.  We assume $\sqrt{s}=510$ GeV, suitable for the 2013 run of RHIC.  Following the single-spin asymmetry analysis of Ref.~\cite{Adam:2018bam} we impose the following cut on the transverse momentum of the final-state lepton: $25 < p_T^e < 50$ GeV.  We use the NNPDFpol1.1 NLO parton distribution function extraction for polarized PDFs~\cite{Nocera:2014gqa}, while for unpolarized PDFs we use NNPDF 3.0 at NNLO~\cite{Ball:2014uwa}.  In principle our polarized cross section should be combined with a NNLO extraction of polarized PDFs, but as discussed in the introduction no such set currently exists.  For the central values of our predictions we assume that the renormalization and factorization scales are equal to $\mu_R = \mu_F=M_W$. To estimate the residual 
theoretical error we vary the scales separately around this central value by a factor of two in each direction, subject to the constraint 
$1/2 \leq \mu_R/\mu_F \leq 2$.  To calculate the scale uncertainties of $A_L$, we treat the numerator and denominator as fully correlated. Since experimentally the same events appear in both the numerator and denominator of the asymmetry, only combined in a different way, we believe that it is reasonable to assume correlated scales in the numerator and denominator of the theoretical prediction.  If the scales are instead treated as uncorrelated, the scale uncertainty increases slightly but is still much smaller than the PDF uncertainty.

\begin{figure}[htbp]
\centering
\includegraphics[width=0.45\textwidth]{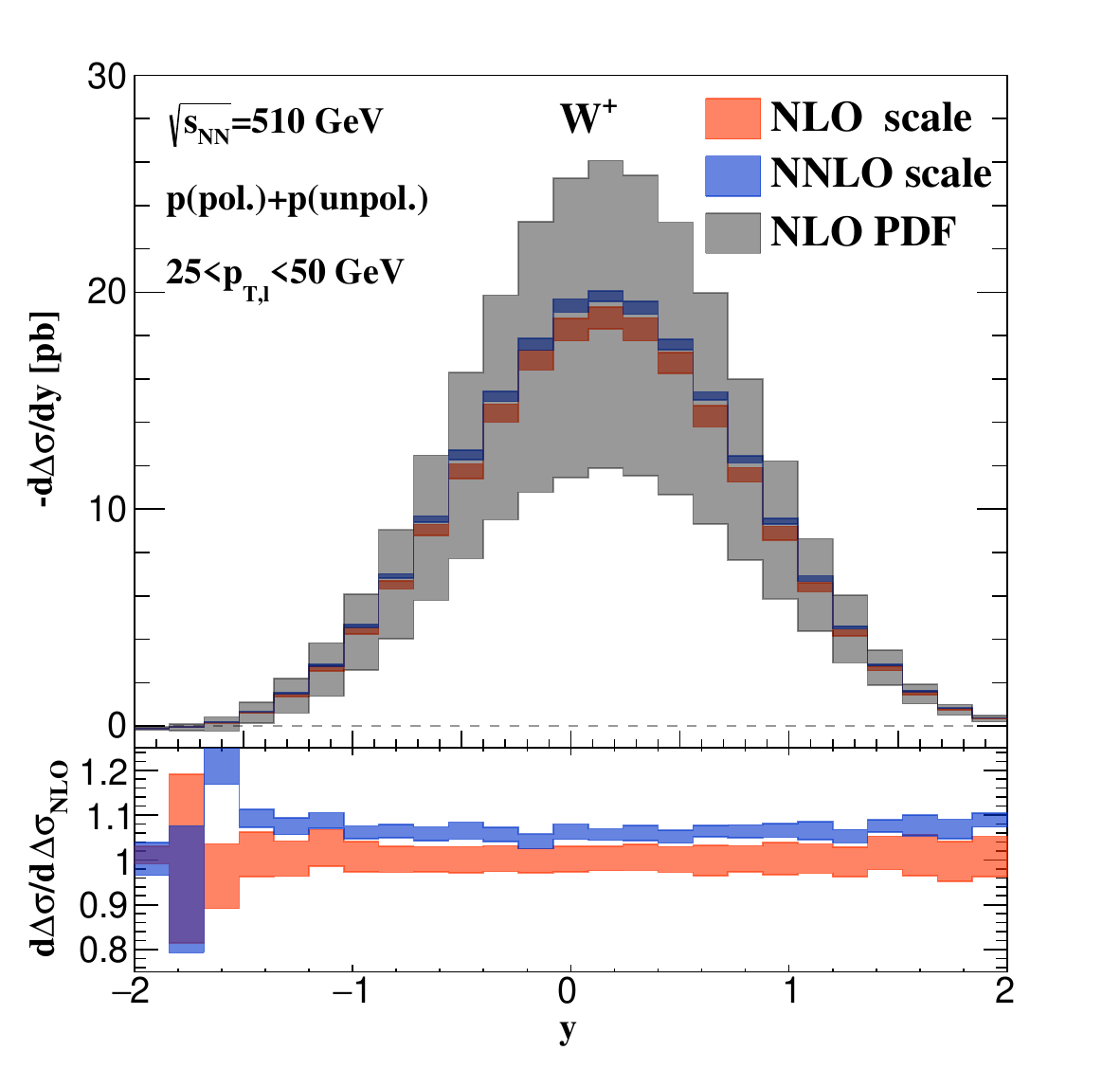}
\vspace{-0.5cm}
\caption{NLO and NNLO results for the lepton rapidity distribution in polarized proton-proton collisions. The lower inset shows the ratio of NNLO over NLO cross sections.  The colored bands denote the scale uncertainties at each perturbative order, while the gray band denotes the PDF uncertainty. Since the polarized cross section is negative we have inserted an overall minus sign in the y-axis.
\label{fig:leprap}}
\end{figure}

\begin{figure}[htbp]
\centering
\includegraphics[width=0.45\textwidth]{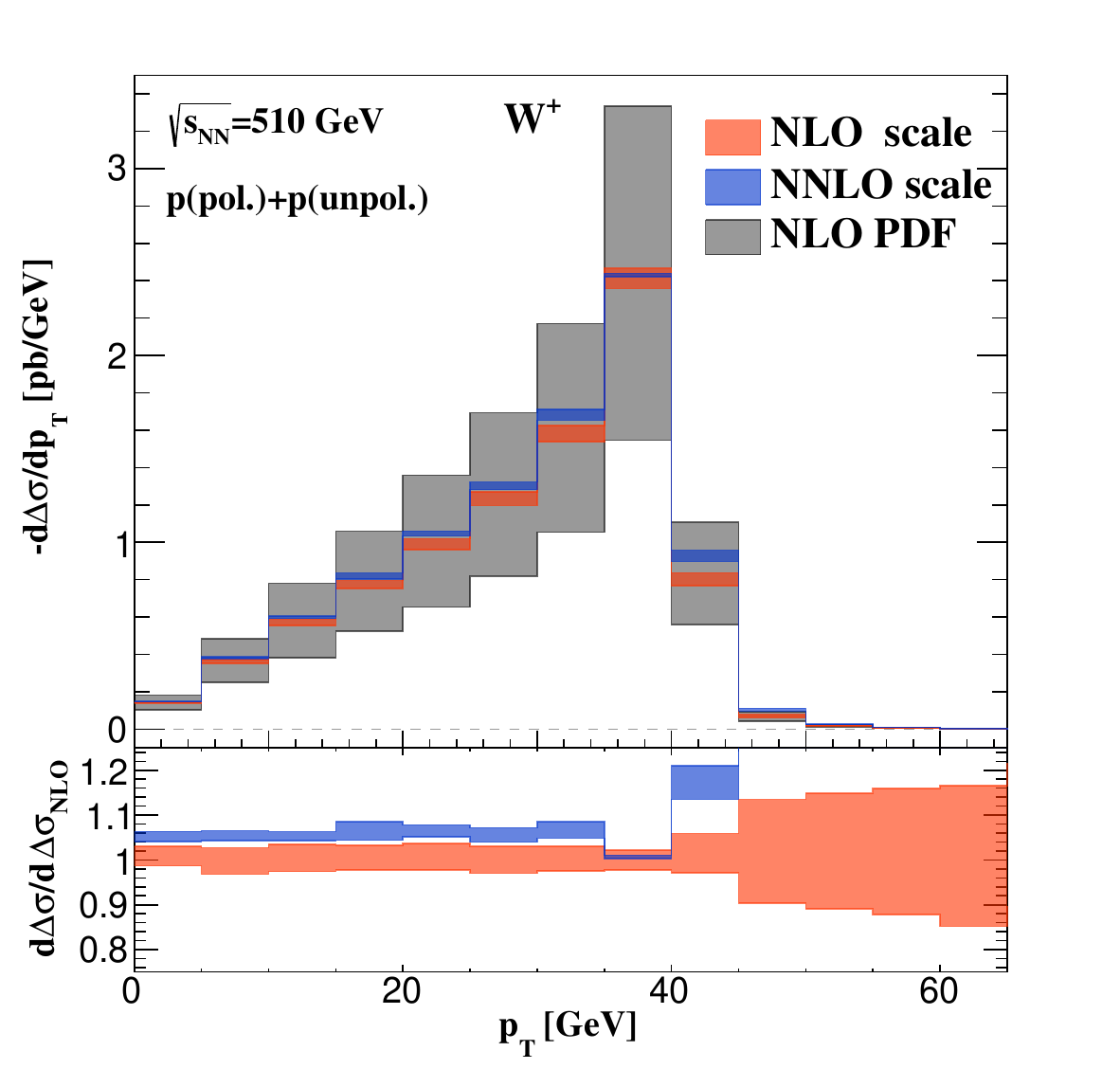}
\vspace{-0.5cm}
\caption{NLO and NNLO results for the lepton transverse momentum distribution in polarized proton-proton collisions. The lower inset shows the ratio of NNLO over NLO cross sections. The colored bands denote the scale uncertainties at each perturbative order, while the gray band denotes the PDF uncertainty. Since the polarized cross section is negative we have inserted an overall minus sign in the y-axis. 
The fiducial cut on the lepton-$p_T$ discussed in the beginning of Section~\ref{sec:numerics} has not been imposed for this distribution.
\label{fig:leppt}}
\end{figure}

We begin by showing results for the polarized cross section $\Delta \sigma$ differential in lepton $p_T$ and rapidity in Figs.~\ref{fig:leprap} and~\ref{fig:leppt} respectively.  Both distributions show an approximately 5\% increase over the NLO prediction upon including the NNLO corrections. The scale uncertainties decrease from $\pm 4\%$ at NLO to $\pm 2\%$ at NNLO over most of phase space.  There are a few kinematic regions where a different pattern of corrections and scale uncertainties is observed, namely for lepton transverse momenta $p_T > 40$ GeV.  This is associated with the Jacobian peak of the $W$-boson.  At LO in QCD perturbation theory there is a restriction $p_T \leq M_W/2$ that is broken minimally only by the finite $W$-width. This restriction is relaxed at NLO, leading to larger corrections and error estimates in this region. We note that the phase-space region beyond the Jacobian peak is only allowed upon the emission of an additional gluon (neglecting the small effect from the $W$-boson width), and are therefore only NLO accurate in the calculation rather than NNLO in the calculation presented here\footnote{Note that additional care is needed for the cross section close to the Jacobian peak, as discussed in Refs.~\cite{Catani:1997xc,Ebert:2020dfc}. If the measurement is restricted to a region very near the peak then soft gluon effects dominate, and large logarithms spoil the convergence of the perturbative QCD expansion. Techniques such as N-jettiness subtraction which introduce a phase-space slicing parameter are not expected to reproduce this region exactly~\cite{Frixione:1997ks}. These issues can be ameliorated by choosing a sufficiently large bin around the peak.}.  With the loose cut 25$<p_{T,l}<$ 50 GeV, an enhancement of the power corrections expected in the presence of fiducial cuts~\cite{Ebert:2019zkb} is not observed in our work.  The PDF uncertainties are calculated by varying the polarized PDFs while with the unpolarized PDFs fixed. The PDF uncertainties dominate over the residual scale uncertainties over all of phase space.  We note that over much of phase space the NNLO result does not lie within the NLO scale variation band. This is similar to that observed for $W$-boson production at the LHC~\cite{Anastasiou:2003yy}, and demonstrates the well-known limitation of relying solely on scale variation to estimate the theoretical uncertainty.

\begin{figure}[htbp]
\centering
\includegraphics[width=0.45\textwidth]{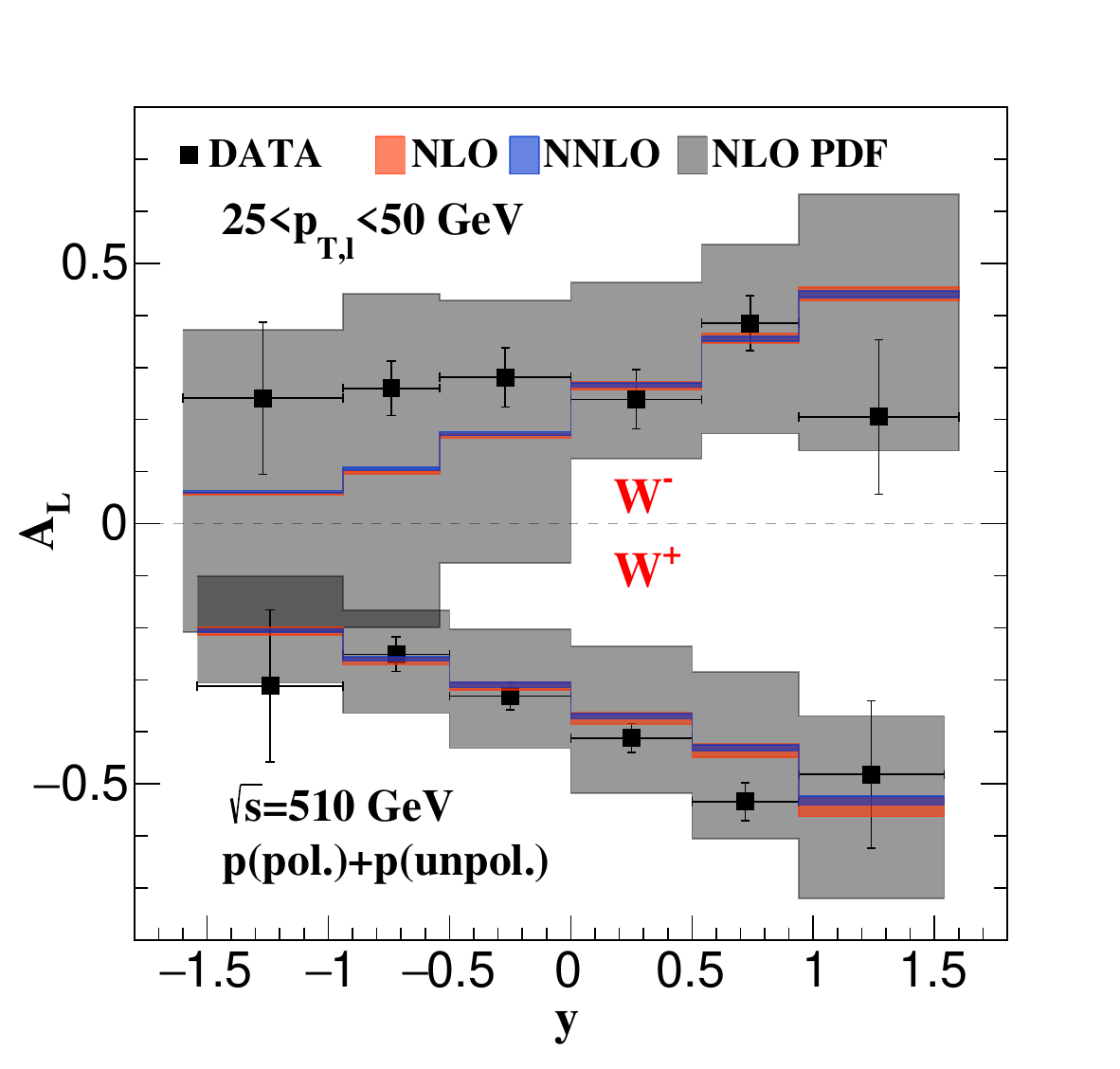}
\vspace{-0.5cm}
\caption{NLO and NNLO predictions for the single-spin asymmetries for $W^-$ production (upper panel) and $W^+$ production (lower panel) compared to experimental data from the STAR collaboration.  Shown also are the residual scale uncertainties at NLO and NNLO as well as the estimated PDF errors. 
\label{fig:AL}}
\end{figure}

We can use our results for the polarized cross section to extend the 
calculation of the single-spin asymmetry $A_L$ to NNLO. We show our result for the asymmetry as a function of the lepton rapidity in Fig.~\ref{fig:AL}.  We compare our predictions at NLO and NNLO to the STAR data of Ref.~\cite{Adam:2018bam}. The theoretical predictions at both NLO and NNLO are in good agreement with the experimental data.  The QCD corrections and scale uncertainties tend to cancel between the numerator and denominator of the asymmetry, leading to small corrections and a greatly reduced scale uncertainty.  The PDF error dominates over the other theoretical uncertainties and the experimental uncertainty. Our results show that the prediction for the asymmetry is completely stable under perturbative QCD corrections. 

\section{Conclusions} \label{sec:conc}

We have presented a calculation of the NNLO perturbative corrections to $W^{\pm}$ production in polarized proton-proton collisions.  Measurements of this process provide the first evidence for flavor asymmetries in the polarized sea-quark distributions, and are important for extending our knowledge of  polarized proton structure.  We find that the NNLO corrections reduce the residual theoretical errors arising from scale dependence.  The single-spin asymmetries from which the polarized sea-quark distributions are determined are extremely stable against NNLO QCD corrections.  Our calculation is a first step toward extending the perturbative order of polarized PDF extractions to the NNLO level by providing the requisite hard-scattering cross sections to this same precision.

\section{Acknowledgments}

R.~B. is supported by the DOE contract DE-AC02-06CH11357. H.T.~L. is supported by the DOE contract DE-AC02-06CH11357 and the NSF grant NSF-1740142. F.~P. is supported by the DOE grants DE-FG02-91ER40684 and DE-AC02-06CH11357.

\bibliography{bib}

\begin{thebibliography}{52}
\expandafter\ifx\csname natexlab\endcsname\relax\def\natexlab#1{#1}\fi
\expandafter\ifx\csname bibnamefont\endcsname\relax
  \def\bibnamefont#1{#1}\fi
\expandafter\ifx\csname bibfnamefont\endcsname\relax
  \def\bibfnamefont#1{#1}\fi
\expandafter\ifx\csname citenamefont\endcsname\relax
  \def\citenamefont#1{#1}\fi
\expandafter\ifx\csname url\endcsname\relax
  \def\url#1{\texttt{#1}}\fi
\expandafter\ifx\csname urlprefix\endcsname\relax\def\urlprefix{URL }\fi
\providecommand{\bibinfo}[2]{#2}
\providecommand{\eprint}[2][]{\url{#2}}

\bibitem[{\citenamefont{Accardi et~al.}(2016)}]{Accardi:2012qut}
\bibinfo{author}{\bibfnamefont{A.}~\bibnamefont{Accardi}} \bibnamefont{et~al.},
  \bibinfo{journal}{Eur. Phys. J. A} \textbf{\bibinfo{volume}{52}},
  \bibinfo{pages}{268} (\bibinfo{year}{2016}), \eprint{1212.1701}.

\bibitem[{\citenamefont{Aschenauer et~al.}(2014)}]{Aschenauer:2014cki}
\bibinfo{author}{\bibfnamefont{E.}~\bibnamefont{Aschenauer}}
  \bibnamefont{et~al.} (\bibinfo{year}{2014}), \eprint{1409.1633}.

\bibitem[{\citenamefont{de~Florian et~al.}(2014)\citenamefont{de~Florian,
  Sassot, Stratmann, and Vogelsang}}]{deFlorian:2014yva}
\bibinfo{author}{\bibfnamefont{D.}~\bibnamefont{de~Florian}},
  \bibinfo{author}{\bibfnamefont{R.}~\bibnamefont{Sassot}},
  \bibinfo{author}{\bibfnamefont{M.}~\bibnamefont{Stratmann}},
  \bibnamefont{and}
  \bibinfo{author}{\bibfnamefont{W.}~\bibnamefont{Vogelsang}},
  \bibinfo{journal}{Phys. Rev. Lett.} \textbf{\bibinfo{volume}{113}},
  \bibinfo{pages}{012001} (\bibinfo{year}{2014}), \eprint{1404.4293}.

\bibitem[{\citenamefont{Bourrely and Soffer}(1993)}]{Bourrely:1993dd}
\bibinfo{author}{\bibfnamefont{C.}~\bibnamefont{Bourrely}} \bibnamefont{and}
  \bibinfo{author}{\bibfnamefont{J.}~\bibnamefont{Soffer}},
  \bibinfo{journal}{Phys. Lett. B} \textbf{\bibinfo{volume}{314}},
  \bibinfo{pages}{132} (\bibinfo{year}{1993}).

\bibitem[{\citenamefont{Bourrely and Soffer}(1994)}]{Bourrely:1994sc}
\bibinfo{author}{\bibfnamefont{C.}~\bibnamefont{Bourrely}} \bibnamefont{and}
  \bibinfo{author}{\bibfnamefont{J.}~\bibnamefont{Soffer}},
  \bibinfo{journal}{Nucl. Phys. B} \textbf{\bibinfo{volume}{423}},
  \bibinfo{pages}{329} (\bibinfo{year}{1994}), \eprint{hep-ph/9405250}.

\bibitem[{\citenamefont{Ethier and Nocera}(2020)}]{Ethier:2020way}
\bibinfo{author}{\bibfnamefont{J.~J.} \bibnamefont{Ethier}} \bibnamefont{and}
  \bibinfo{author}{\bibfnamefont{E.~R.} \bibnamefont{Nocera}},
  \bibinfo{journal}{Ann. Rev. Nucl. Part. Sci.} \textbf{\bibinfo{volume}{70}},
  \bibinfo{pages}{43} (\bibinfo{year}{2020}), \eprint{2001.07722}.

\bibitem[{\citenamefont{Aggarwal et~al.}(2011)}]{Aggarwal:2010vc}
\bibinfo{author}{\bibfnamefont{M.}~\bibnamefont{Aggarwal}} \bibnamefont{et~al.}
  (\bibinfo{collaboration}{STAR}), \bibinfo{journal}{Phys. Rev. Lett.}
  \textbf{\bibinfo{volume}{106}}, \bibinfo{pages}{062002}
  (\bibinfo{year}{2011}), \eprint{1009.0326}.

\bibitem[{\citenamefont{Adamczyk et~al.}(2014)}]{Adamczyk:2014xyw}
\bibinfo{author}{\bibfnamefont{L.}~\bibnamefont{Adamczyk}} \bibnamefont{et~al.}
  (\bibinfo{collaboration}{STAR}), \bibinfo{journal}{Phys. Rev. Lett.}
  \textbf{\bibinfo{volume}{113}}, \bibinfo{pages}{072301}
  (\bibinfo{year}{2014}), \eprint{1404.6880}.

\bibitem[{\citenamefont{Adare et~al.}(2018)}]{Adare:2018csm}
\bibinfo{author}{\bibfnamefont{A.}~\bibnamefont{Adare}} \bibnamefont{et~al.}
  (\bibinfo{collaboration}{PHENIX}), \bibinfo{journal}{Phys. Rev. D}
  \textbf{\bibinfo{volume}{98}}, \bibinfo{pages}{032007}
  (\bibinfo{year}{2018}), \eprint{1804.04181}.

\bibitem[{\citenamefont{Adam et~al.}(2019)}]{Adam:2018bam}
\bibinfo{author}{\bibfnamefont{J.}~\bibnamefont{Adam}} \bibnamefont{et~al.}
  (\bibinfo{collaboration}{STAR}), \bibinfo{journal}{Phys. Rev. D}
  \textbf{\bibinfo{volume}{99}}, \bibinfo{pages}{051102}
  (\bibinfo{year}{2019}), \eprint{1812.04817}.

\bibitem[{\citenamefont{Harland-Lang et~al.}(2018)\citenamefont{Harland-Lang,
  Martin, and Thorne}}]{Harland-Lang:2017ytb}
\bibinfo{author}{\bibfnamefont{L.}~\bibnamefont{Harland-Lang}},
  \bibinfo{author}{\bibfnamefont{A.}~\bibnamefont{Martin}}, \bibnamefont{and}
  \bibinfo{author}{\bibfnamefont{R.}~\bibnamefont{Thorne}},
  \bibinfo{journal}{Eur. Phys. J. C} \textbf{\bibinfo{volume}{78}},
  \bibinfo{pages}{248} (\bibinfo{year}{2018}), \eprint{1711.05757}.

\bibitem[{\citenamefont{Ball et~al.}(2017)}]{Ball:2017nwa}
\bibinfo{author}{\bibfnamefont{R.~D.} \bibnamefont{Ball}} \bibnamefont{et~al.}
  (\bibinfo{collaboration}{NNPDF}), \bibinfo{journal}{Eur. Phys. J. C}
  \textbf{\bibinfo{volume}{77}}, \bibinfo{pages}{663} (\bibinfo{year}{2017}),
  \eprint{1706.00428}.

\bibitem[{\citenamefont{Alekhin et~al.}(2017)\citenamefont{Alekhin, Bl\"umlein,
  Moch, and Placakyte}}]{Alekhin:2017kpj}
\bibinfo{author}{\bibfnamefont{S.}~\bibnamefont{Alekhin}},
  \bibinfo{author}{\bibfnamefont{J.}~\bibnamefont{Bl\"umlein}},
  \bibinfo{author}{\bibfnamefont{S.}~\bibnamefont{Moch}}, \bibnamefont{and}
  \bibinfo{author}{\bibfnamefont{R.}~\bibnamefont{Placakyte}},
  \bibinfo{journal}{Phys. Rev. D} \textbf{\bibinfo{volume}{96}},
  \bibinfo{pages}{014011} (\bibinfo{year}{2017}), \eprint{1701.05838}.

\bibitem[{\citenamefont{Hou et~al.}(2019)}]{Hou:2019efy}
\bibinfo{author}{\bibfnamefont{T.-J.} \bibnamefont{Hou}} \bibnamefont{et~al.}
  (\bibinfo{year}{2019}), \eprint{1912.10053}.

\bibitem[{\citenamefont{Butterworth et~al.}(2016)}]{Butterworth:2015oua}
\bibinfo{author}{\bibfnamefont{J.}~\bibnamefont{Butterworth}}
  \bibnamefont{et~al.}, \bibinfo{journal}{J. Phys. G}
  \textbf{\bibinfo{volume}{43}}, \bibinfo{pages}{023001}
  (\bibinfo{year}{2016}), \eprint{1510.03865}.

\bibitem[{\citenamefont{Nocera et~al.}(2014)\citenamefont{Nocera, Ball, Forte,
  Ridolfi, and Rojo}}]{Nocera:2014gqa}
\bibinfo{author}{\bibfnamefont{E.~R.} \bibnamefont{Nocera}},
  \bibinfo{author}{\bibfnamefont{R.~D.} \bibnamefont{Ball}},
  \bibinfo{author}{\bibfnamefont{S.}~\bibnamefont{Forte}},
  \bibinfo{author}{\bibfnamefont{G.}~\bibnamefont{Ridolfi}}, \bibnamefont{and}
  \bibinfo{author}{\bibfnamefont{J.}~\bibnamefont{Rojo}}
  (\bibinfo{collaboration}{NNPDF}), \bibinfo{journal}{Nucl. Phys. B}
  \textbf{\bibinfo{volume}{887}}, \bibinfo{pages}{276} (\bibinfo{year}{2014}),
  \eprint{1406.5539}.

\bibitem[{\citenamefont{Ethier et~al.}(2017)\citenamefont{Ethier, Sato, and
  Melnitchouk}}]{Ethier:2017zbq}
\bibinfo{author}{\bibfnamefont{J.}~\bibnamefont{Ethier}},
  \bibinfo{author}{\bibfnamefont{N.}~\bibnamefont{Sato}}, \bibnamefont{and}
  \bibinfo{author}{\bibfnamefont{W.}~\bibnamefont{Melnitchouk}},
  \bibinfo{journal}{Phys. Rev. Lett.} \textbf{\bibinfo{volume}{119}},
  \bibinfo{pages}{132001} (\bibinfo{year}{2017}), \eprint{1705.05889}.

\bibitem[{\citenamefont{Nadolsky and
  Yuan}(2003{\natexlab{a}})}]{Nadolsky:2003fz}
\bibinfo{author}{\bibfnamefont{P.~M.} \bibnamefont{Nadolsky}} \bibnamefont{and}
  \bibinfo{author}{\bibfnamefont{C.~P.} \bibnamefont{Yuan}},
  \bibinfo{journal}{Nucl. Phys. B} \textbf{\bibinfo{volume}{666}},
  \bibinfo{pages}{3} (\bibinfo{year}{2003}{\natexlab{a}}),
  \eprint{hep-ph/0304001}.

\bibitem[{\citenamefont{Nadolsky and
  Yuan}(2003{\natexlab{b}})}]{Nadolsky:2003ga}
\bibinfo{author}{\bibfnamefont{P.~M.} \bibnamefont{Nadolsky}} \bibnamefont{and}
  \bibinfo{author}{\bibfnamefont{C.~P.} \bibnamefont{Yuan}},
  \bibinfo{journal}{Nucl. Phys. B} \textbf{\bibinfo{volume}{666}},
  \bibinfo{pages}{31} (\bibinfo{year}{2003}{\natexlab{b}}),
  \eprint{hep-ph/0304002}.

\bibitem[{\citenamefont{de~Florian and Vogelsang}(2010)}]{deFlorian:2010aa}
\bibinfo{author}{\bibfnamefont{D.}~\bibnamefont{de~Florian}} \bibnamefont{and}
  \bibinfo{author}{\bibfnamefont{W.}~\bibnamefont{Vogelsang}},
  \bibinfo{journal}{Phys. Rev. D} \textbf{\bibinfo{volume}{81}},
  \bibinfo{pages}{094020} (\bibinfo{year}{2010}), \eprint{1003.4533}.

\bibitem[{\citenamefont{Ringer and Vogelsang}(2015)}]{Ringer:2015oaa}
\bibinfo{author}{\bibfnamefont{F.}~\bibnamefont{Ringer}} \bibnamefont{and}
  \bibinfo{author}{\bibfnamefont{W.}~\bibnamefont{Vogelsang}},
  \bibinfo{journal}{Phys. Rev. D} \textbf{\bibinfo{volume}{91}},
  \bibinfo{pages}{094033} (\bibinfo{year}{2015}), \eprint{1503.07052}.

\bibitem[{\citenamefont{Boughezal et~al.}(2015)\citenamefont{Boughezal, Focke,
  Liu, and Petriello}}]{Boughezal:2015dva}
\bibinfo{author}{\bibfnamefont{R.}~\bibnamefont{Boughezal}},
  \bibinfo{author}{\bibfnamefont{C.}~\bibnamefont{Focke}},
  \bibinfo{author}{\bibfnamefont{X.}~\bibnamefont{Liu}}, \bibnamefont{and}
  \bibinfo{author}{\bibfnamefont{F.}~\bibnamefont{Petriello}},
  \bibinfo{journal}{Phys. Rev. Lett.} \textbf{\bibinfo{volume}{115}},
  \bibinfo{pages}{062002} (\bibinfo{year}{2015}), \eprint{1504.02131}.

\bibitem[{\citenamefont{Gaunt et~al.}(2015)\citenamefont{Gaunt, Stahlhofen,
  Tackmann, and Walsh}}]{Gaunt:2015pea}
\bibinfo{author}{\bibfnamefont{J.}~\bibnamefont{Gaunt}},
  \bibinfo{author}{\bibfnamefont{M.}~\bibnamefont{Stahlhofen}},
  \bibinfo{author}{\bibfnamefont{F.~J.} \bibnamefont{Tackmann}},
  \bibnamefont{and} \bibinfo{author}{\bibfnamefont{J.~R.} \bibnamefont{Walsh}},
  \bibinfo{journal}{JHEP} \textbf{\bibinfo{volume}{09}}, \bibinfo{pages}{058}
  (\bibinfo{year}{2015}), \eprint{1505.04794}.

\bibitem[{\citenamefont{Stewart
  et~al.}(2010{\natexlab{a}})\citenamefont{Stewart, Tackmann, and
  Waalewijn}}]{Stewart:2010tn}
\bibinfo{author}{\bibfnamefont{I.~W.} \bibnamefont{Stewart}},
  \bibinfo{author}{\bibfnamefont{F.~J.} \bibnamefont{Tackmann}},
  \bibnamefont{and} \bibinfo{author}{\bibfnamefont{W.~J.}
  \bibnamefont{Waalewijn}}, \bibinfo{journal}{Phys. Rev. Lett.}
  \textbf{\bibinfo{volume}{105}}, \bibinfo{pages}{092002}
  (\bibinfo{year}{2010}{\natexlab{a}}), \eprint{1004.2489}.

\bibitem[{\citenamefont{Stewart
  et~al.}(2010{\natexlab{b}})\citenamefont{Stewart, Tackmann, and
  Waalewijn}}]{Stewart:2009yx}
\bibinfo{author}{\bibfnamefont{I.~W.} \bibnamefont{Stewart}},
  \bibinfo{author}{\bibfnamefont{F.~J.} \bibnamefont{Tackmann}},
  \bibnamefont{and} \bibinfo{author}{\bibfnamefont{W.~J.}
  \bibnamefont{Waalewijn}}, \bibinfo{journal}{Phys. Rev. D}
  \textbf{\bibinfo{volume}{81}}, \bibinfo{pages}{094035}
  (\bibinfo{year}{2010}{\natexlab{b}}), \eprint{0910.0467}.

\bibitem[{\citenamefont{Stewart et~al.}(2011)\citenamefont{Stewart, Tackmann,
  and Waalewijn}}]{Stewart:2010pd}
\bibinfo{author}{\bibfnamefont{I.~W.} \bibnamefont{Stewart}},
  \bibinfo{author}{\bibfnamefont{F.~J.} \bibnamefont{Tackmann}},
  \bibnamefont{and} \bibinfo{author}{\bibfnamefont{W.~J.}
  \bibnamefont{Waalewijn}}, \bibinfo{journal}{Phys. Rev. Lett.}
  \textbf{\bibinfo{volume}{106}}, \bibinfo{pages}{032001}
  (\bibinfo{year}{2011}), \eprint{1005.4060}.

\bibitem[{\citenamefont{Moult et~al.}(2017)\citenamefont{Moult, Rothen,
  Stewart, Tackmann, and Zhu}}]{Moult:2016fqy}
\bibinfo{author}{\bibfnamefont{I.}~\bibnamefont{Moult}},
  \bibinfo{author}{\bibfnamefont{L.}~\bibnamefont{Rothen}},
  \bibinfo{author}{\bibfnamefont{I.~W.} \bibnamefont{Stewart}},
  \bibinfo{author}{\bibfnamefont{F.~J.} \bibnamefont{Tackmann}},
  \bibnamefont{and} \bibinfo{author}{\bibfnamefont{H.~X.} \bibnamefont{Zhu}},
  \bibinfo{journal}{Phys. Rev. D} \textbf{\bibinfo{volume}{95}},
  \bibinfo{pages}{074023} (\bibinfo{year}{2017}), \eprint{1612.00450}.

\bibitem[{\citenamefont{Boughezal
  et~al.}(2017{\natexlab{a}})\citenamefont{Boughezal, Liu, and
  Petriello}}]{Boughezal:2016zws}
\bibinfo{author}{\bibfnamefont{R.}~\bibnamefont{Boughezal}},
  \bibinfo{author}{\bibfnamefont{X.}~\bibnamefont{Liu}}, \bibnamefont{and}
  \bibinfo{author}{\bibfnamefont{F.}~\bibnamefont{Petriello}},
  \bibinfo{journal}{JHEP} \textbf{\bibinfo{volume}{03}}, \bibinfo{pages}{160}
  (\bibinfo{year}{2017}{\natexlab{a}}), \eprint{1612.02911}.

\bibitem[{\citenamefont{Moult et~al.}(2018)\citenamefont{Moult, Rothen,
  Stewart, Tackmann, and Zhu}}]{Moult:2017jsg}
\bibinfo{author}{\bibfnamefont{I.}~\bibnamefont{Moult}},
  \bibinfo{author}{\bibfnamefont{L.}~\bibnamefont{Rothen}},
  \bibinfo{author}{\bibfnamefont{I.~W.} \bibnamefont{Stewart}},
  \bibinfo{author}{\bibfnamefont{F.~J.} \bibnamefont{Tackmann}},
  \bibnamefont{and} \bibinfo{author}{\bibfnamefont{H.~X.} \bibnamefont{Zhu}},
  \bibinfo{journal}{Phys. Rev. D} \textbf{\bibinfo{volume}{97}},
  \bibinfo{pages}{014013} (\bibinfo{year}{2018}), \eprint{1710.03227}.

\bibitem[{\citenamefont{Boughezal et~al.}(2018)\citenamefont{Boughezal,
  Isgr\`o, and Petriello}}]{Boughezal:2018mvf}
\bibinfo{author}{\bibfnamefont{R.}~\bibnamefont{Boughezal}},
  \bibinfo{author}{\bibfnamefont{A.}~\bibnamefont{Isgr\`o}}, \bibnamefont{and}
  \bibinfo{author}{\bibfnamefont{F.}~\bibnamefont{Petriello}},
  \bibinfo{journal}{Phys. Rev. D} \textbf{\bibinfo{volume}{97}},
  \bibinfo{pages}{076006} (\bibinfo{year}{2018}), \eprint{1802.00456}.

\bibitem[{\citenamefont{Ebert et~al.}(2018)\citenamefont{Ebert, Moult, Stewart,
  Tackmann, Vita, and Zhu}}]{Ebert:2018lzn}
\bibinfo{author}{\bibfnamefont{M.~A.} \bibnamefont{Ebert}},
  \bibinfo{author}{\bibfnamefont{I.}~\bibnamefont{Moult}},
  \bibinfo{author}{\bibfnamefont{I.~W.} \bibnamefont{Stewart}},
  \bibinfo{author}{\bibfnamefont{F.~J.} \bibnamefont{Tackmann}},
  \bibinfo{author}{\bibfnamefont{G.}~\bibnamefont{Vita}}, \bibnamefont{and}
  \bibinfo{author}{\bibfnamefont{H.~X.} \bibnamefont{Zhu}},
  \bibinfo{journal}{JHEP} \textbf{\bibinfo{volume}{12}}, \bibinfo{pages}{084}
  (\bibinfo{year}{2018}), \eprint{1807.10764}.

\bibitem[{\citenamefont{Boughezal et~al.}(2020)\citenamefont{Boughezal,
  Isgr\`o, and Petriello}}]{Boughezal:2019ggi}
\bibinfo{author}{\bibfnamefont{R.}~\bibnamefont{Boughezal}},
  \bibinfo{author}{\bibfnamefont{A.}~\bibnamefont{Isgr\`o}}, \bibnamefont{and}
  \bibinfo{author}{\bibfnamefont{F.}~\bibnamefont{Petriello}},
  \bibinfo{journal}{Phys. Rev. D} \textbf{\bibinfo{volume}{101}},
  \bibinfo{pages}{016005} (\bibinfo{year}{2020}), \eprint{1907.12213}.

\bibitem[{\citenamefont{Stewart
  et~al.}(2010{\natexlab{c}})\citenamefont{Stewart, Tackmann, and
  Waalewijn}}]{Stewart:2010qs}
\bibinfo{author}{\bibfnamefont{I.~W.} \bibnamefont{Stewart}},
  \bibinfo{author}{\bibfnamefont{F.~J.} \bibnamefont{Tackmann}},
  \bibnamefont{and} \bibinfo{author}{\bibfnamefont{W.~J.}
  \bibnamefont{Waalewijn}}, \bibinfo{journal}{JHEP}
  \textbf{\bibinfo{volume}{09}}, \bibinfo{pages}{005}
  (\bibinfo{year}{2010}{\natexlab{c}}), \eprint{1002.2213}.

\bibitem[{\citenamefont{Gaunt et~al.}(2014)\citenamefont{Gaunt, Stahlhofen, and
  Tackmann}}]{Gaunt:2014xga}
\bibinfo{author}{\bibfnamefont{J.~R.} \bibnamefont{Gaunt}},
  \bibinfo{author}{\bibfnamefont{M.}~\bibnamefont{Stahlhofen}},
  \bibnamefont{and} \bibinfo{author}{\bibfnamefont{F.~J.}
  \bibnamefont{Tackmann}}, \bibinfo{journal}{JHEP}
  \textbf{\bibinfo{volume}{04}}, \bibinfo{pages}{113} (\bibinfo{year}{2014}),
  \eprint{1401.5478}.

\bibitem[{\citenamefont{Kelley et~al.}(2011)\citenamefont{Kelley, Schwartz,
  Schabinger, and Zhu}}]{Kelley:2011ng}
\bibinfo{author}{\bibfnamefont{R.}~\bibnamefont{Kelley}},
  \bibinfo{author}{\bibfnamefont{M.~D.} \bibnamefont{Schwartz}},
  \bibinfo{author}{\bibfnamefont{R.~M.} \bibnamefont{Schabinger}},
  \bibnamefont{and} \bibinfo{author}{\bibfnamefont{H.~X.} \bibnamefont{Zhu}},
  \bibinfo{journal}{Phys. Rev. D} \textbf{\bibinfo{volume}{84}},
  \bibinfo{pages}{045022} (\bibinfo{year}{2011}), \eprint{1105.3676}.

\bibitem[{\citenamefont{Monni et~al.}(2011)\citenamefont{Monni, Gehrmann, and
  Luisoni}}]{Monni:2011gb}
\bibinfo{author}{\bibfnamefont{P.~F.} \bibnamefont{Monni}},
  \bibinfo{author}{\bibfnamefont{T.}~\bibnamefont{Gehrmann}}, \bibnamefont{and}
  \bibinfo{author}{\bibfnamefont{G.}~\bibnamefont{Luisoni}},
  \bibinfo{journal}{JHEP} \textbf{\bibinfo{volume}{08}}, \bibinfo{pages}{010}
  (\bibinfo{year}{2011}), \eprint{1105.4560}.

\bibitem[{\citenamefont{Hornig et~al.}(2011)\citenamefont{Hornig, Lee, Stewart,
  Walsh, and Zuberi}}]{Hornig:2011iu}
\bibinfo{author}{\bibfnamefont{A.}~\bibnamefont{Hornig}},
  \bibinfo{author}{\bibfnamefont{C.}~\bibnamefont{Lee}},
  \bibinfo{author}{\bibfnamefont{I.~W.} \bibnamefont{Stewart}},
  \bibinfo{author}{\bibfnamefont{J.~R.} \bibnamefont{Walsh}}, \bibnamefont{and}
  \bibinfo{author}{\bibfnamefont{S.}~\bibnamefont{Zuberi}},
  \bibinfo{journal}{JHEP} \textbf{\bibinfo{volume}{08}}, \bibinfo{pages}{054}
  (\bibinfo{year}{2011}), \bibinfo{note}{[Erratum: JHEP 10, 101 (2017)]},
  \eprint{1105.4628}.


\bibitem[{\citenamefont{Catani and Seymour}(1997)}]{Catani:1996vz}
\bibinfo{author}{\bibfnamefont{S.}~\bibnamefont{Catani}} \bibnamefont{and}
  \bibinfo{author}{\bibfnamefont{M.}~\bibnamefont{Seymour}},
  \bibinfo{journal}{Nucl. Phys. B} \textbf{\bibinfo{volume}{485}},
  \bibinfo{pages}{291} (\bibinfo{year}{1997}), \bibinfo{note}{[Erratum:
  Nucl.Phys.B 510, 503--504 (1998)]}, \eprint{hep-ph/9605323}.

\bibitem[{\citenamefont{Boughezal
  et~al.}(2017{\natexlab{b}})\citenamefont{Boughezal, Campbell, Ellis, Focke,
  Giele, Liu, Petriello, and Williams}}]{Boughezal:2016wmq}
\bibinfo{author}{\bibfnamefont{R.}~\bibnamefont{Boughezal}},
  \bibinfo{author}{\bibfnamefont{J.~M.} \bibnamefont{Campbell}},
  \bibinfo{author}{\bibfnamefont{R.~K.} \bibnamefont{Ellis}},
  \bibinfo{author}{\bibfnamefont{C.}~\bibnamefont{Focke}},
  \bibinfo{author}{\bibfnamefont{W.}~\bibnamefont{Giele}},
  \bibinfo{author}{\bibfnamefont{X.}~\bibnamefont{Liu}},
  \bibinfo{author}{\bibfnamefont{F.}~\bibnamefont{Petriello}},
  \bibnamefont{and} \bibinfo{author}{\bibfnamefont{C.}~\bibnamefont{Williams}},
  \bibinfo{journal}{Eur. Phys. J. C} \textbf{\bibinfo{volume}{77}},
  \bibinfo{pages}{7} (\bibinfo{year}{2017}{\natexlab{b}}), \eprint{1605.08011}.

\bibitem[{\citenamefont{Baikov et~al.}(2009)\citenamefont{Baikov, Chetyrkin,
  Smirnov, Smirnov, and Steinhauser}}]{Baikov:2009bg}
\bibinfo{author}{\bibfnamefont{P.}~\bibnamefont{Baikov}},
  \bibinfo{author}{\bibfnamefont{K.}~\bibnamefont{Chetyrkin}},
  \bibinfo{author}{\bibfnamefont{A.}~\bibnamefont{Smirnov}},
  \bibinfo{author}{\bibfnamefont{V.}~\bibnamefont{Smirnov}}, \bibnamefont{and}
  \bibinfo{author}{\bibfnamefont{M.}~\bibnamefont{Steinhauser}},
  \bibinfo{journal}{Phys. Rev. Lett.} \textbf{\bibinfo{volume}{102}},
  \bibinfo{pages}{212002} (\bibinfo{year}{2009}), \eprint{0902.3519}.

\bibitem[{\citenamefont{Gehrmann et~al.}(2010)\citenamefont{Gehrmann, Glover,
  Huber, Ikizlerli, and Studerus}}]{Gehrmann:2010ue}
\bibinfo{author}{\bibfnamefont{T.}~\bibnamefont{Gehrmann}},
  \bibinfo{author}{\bibfnamefont{E.}~\bibnamefont{Glover}},
  \bibinfo{author}{\bibfnamefont{T.}~\bibnamefont{Huber}},
  \bibinfo{author}{\bibfnamefont{N.}~\bibnamefont{Ikizlerli}},
  \bibnamefont{and} \bibinfo{author}{\bibfnamefont{C.}~\bibnamefont{Studerus}},
  \bibinfo{journal}{JHEP} \textbf{\bibinfo{volume}{06}}, \bibinfo{pages}{094}
  (\bibinfo{year}{2010}), \eprint{1004.3653}.

\bibitem[{\citenamefont{Boughezal
  et~al.}(2017{\natexlab{c}})\citenamefont{Boughezal, Petriello, Schubert, and
  Xing}}]{Boughezal:2017tdd}
\bibinfo{author}{\bibfnamefont{R.}~\bibnamefont{Boughezal}},
  \bibinfo{author}{\bibfnamefont{F.}~\bibnamefont{Petriello}},
  \bibinfo{author}{\bibfnamefont{U.}~\bibnamefont{Schubert}}, \bibnamefont{and}
  \bibinfo{author}{\bibfnamefont{H.}~\bibnamefont{Xing}},
  \bibinfo{journal}{Phys. Rev. D} \textbf{\bibinfo{volume}{96}},
  \bibinfo{pages}{034001} (\bibinfo{year}{2017}{\natexlab{c}}),
  \eprint{1704.05457}.

\bibitem[{\citenamefont{Campbell and Ellis}(2010)}]{Campbell:2010ff}
\bibinfo{author}{\bibfnamefont{J.~M.} \bibnamefont{Campbell}} \bibnamefont{and}
  \bibinfo{author}{\bibfnamefont{R.}~\bibnamefont{Ellis}},
  \bibinfo{journal}{Nucl. Phys. B Proc. Suppl.}
  \textbf{\bibinfo{volume}{205-206}}, \bibinfo{pages}{10}
  (\bibinfo{year}{2010}), \eprint{1007.3492}.

\bibitem[{\citenamefont{Czakon et~al.}(2009)\citenamefont{Czakon, Papadopoulos,
  and Worek}}]{Czakon:2009ss}
\bibinfo{author}{\bibfnamefont{M.}~\bibnamefont{Czakon}},
  \bibinfo{author}{\bibfnamefont{C.}~\bibnamefont{Papadopoulos}},
  \bibnamefont{and} \bibinfo{author}{\bibfnamefont{M.}~\bibnamefont{Worek}},
  \bibinfo{journal}{JHEP} \textbf{\bibinfo{volume}{08}}, \bibinfo{pages}{085}
  (\bibinfo{year}{2009}), \eprint{0905.0883}.

\bibitem[{\citenamefont{Vogelsang}(1996)}]{Vogelsang:1996im}
\bibinfo{author}{\bibfnamefont{W.}~\bibnamefont{Vogelsang}},
  \bibinfo{journal}{Nucl. Phys. B} \textbf{\bibinfo{volume}{475}},
  \bibinfo{pages}{47} (\bibinfo{year}{1996}), \eprint{hep-ph/9603366}.

\bibitem[{\citenamefont{Ball et~al.}(2015)}]{Ball:2014uwa}
\bibinfo{author}{\bibfnamefont{R.~D.} \bibnamefont{Ball}} \bibnamefont{et~al.}
  (\bibinfo{collaboration}{NNPDF}), \bibinfo{journal}{JHEP}
  \textbf{\bibinfo{volume}{04}}, \bibinfo{pages}{040} (\bibinfo{year}{2015}),
  \eprint{1410.8849}.

\bibitem[{\citenamefont{Catani and Webber}(1997)}]{Catani:1997xc}
\bibinfo{author}{\bibfnamefont{S.}~\bibnamefont{Catani}} \bibnamefont{and}
  \bibinfo{author}{\bibfnamefont{B.~R.} \bibnamefont{Webber}},
  \bibinfo{journal}{JHEP} \textbf{\bibinfo{volume}{10}}, \bibinfo{pages}{005}
  (\bibinfo{year}{1997}), \eprint{hep-ph/9710333}.

\bibitem[{\citenamefont{Ebert et~al.}(2020{\natexlab{b}})\citenamefont{Ebert,
  Michel, Stewart, and Tackmann}}]{Ebert:2020dfc}
\bibinfo{author}{\bibfnamefont{M.~A.} \bibnamefont{Ebert}},
  \bibinfo{author}{\bibfnamefont{J.~K.~L.} \bibnamefont{Michel}},
  \bibinfo{author}{\bibfnamefont{I.~W.} \bibnamefont{Stewart}},
  \bibnamefont{and} \bibinfo{author}{\bibfnamefont{F.~J.}
  \bibnamefont{Tackmann}} (\bibinfo{year}{2020}{\natexlab{b}}),
  \eprint{2006.11382}.

\bibitem[{\citenamefont{Frixione and Ridolfi}(1997)}]{Frixione:1997ks}
\bibinfo{author}{\bibfnamefont{S.}~\bibnamefont{Frixione}} \bibnamefont{and}
  \bibinfo{author}{\bibfnamefont{G.}~\bibnamefont{Ridolfi}},
  \bibinfo{journal}{Nucl. Phys. B} \textbf{\bibinfo{volume}{507}},
  \bibinfo{pages}{315} (\bibinfo{year}{1997}), \eprint{hep-ph/9707345}.

\bibitem[{\citenamefont{Ebert and Tackmann}(2020)}]{Ebert:2019zkb}
\bibinfo{author}{\bibfnamefont{M.~A.} \bibnamefont{Ebert}} \bibnamefont{and}
  \bibinfo{author}{\bibfnamefont{F.~J.} \bibnamefont{Tackmann}},
  \bibinfo{journal}{JHEP} \textbf{\bibinfo{volume}{03}}, \bibinfo{pages}{158}
  (\bibinfo{year}{2020}), \eprint{1911.08486}.

\bibitem[{\citenamefont{Anastasiou et~al.}(2003)\citenamefont{Anastasiou,
  Dixon, Melnikov, and Petriello}}]{Anastasiou:2003yy}
\bibinfo{author}{\bibfnamefont{C.}~\bibnamefont{Anastasiou}},
  \bibinfo{author}{\bibfnamefont{L.~J.} \bibnamefont{Dixon}},
  \bibinfo{author}{\bibfnamefont{K.}~\bibnamefont{Melnikov}}, \bibnamefont{and}
  \bibinfo{author}{\bibfnamefont{F.}~\bibnamefont{Petriello}},
  \bibinfo{journal}{Phys. Rev. Lett.} \textbf{\bibinfo{volume}{91}},
  \bibinfo{pages}{182002} (\bibinfo{year}{2003}), \eprint{hep-ph/0306192}.

\end{thebibliography}

\end{document}